
\hrule height0pt
\magnification=\magstep1
\baselineskip 12pt
\parskip=6pt
\parindent=0pt
\vsize=9 truein
\hsize=6.5 truein
\raggedbottom
\def\pp{\parshape 2 0truecm 15truecm 2.0truecm 15.0truecm}
\def\ref #1;#2;#3;#4{\par\pp #1, {\it #2}, {\bf #3}, #4}
\def\book #1;#2;#3{\par\pp #1, {\it #2}, #3}
\def\rep #1;#2;#3{\par\pp #1, #2, #3}

\def\simlt{\lower.5ex\hbox{$\; \buildrel < \over \sim \;$}}
\def\simgt{\lower.5ex\hbox{$\; \buildrel > \over \sim \;$}}
\def\kms{km\ s$^{-1}$}
\def\Mpc{Mpc}
\def\et{{\it et al.\ }}

\hyphenation{aniso-tropies}

\hsize=6.5truein
\vsize=8.3truein


\font\titlefont=cmss17

\centerline{\titlefont Dynamical and Observable Constraints on RAMBOs:}
\vskip 0.16truein
\centerline{\titlefont Robust Associations of Massive Baryonic Objects}

\vskip 0.8truein
\centerline{\titlefont Ben Moore \ and \ Joseph Silk}
\vskip 0.3truein

\centerline{\it Department of Astronomy \& Center for Particle Astrophysics}
\centerline{\it University of California, Berkeley, CA 94720, USA}
\vskip 0.5truein

\centerline {\bf ABSTRACT}
\vskip 8pt
\parindent=36pt
\vskip 0.3truecm

If the halo dark matter consists of faint baryonic stars, then these objects
probably formed at an early epoch within large associations with similar
dynamical properties to globular or open clusters. We use the luminosity
function of globular clusters as a function of galactocentric distance to
provide a
strong constraint on the properties of RAMBOs. We show that at
the solar radius, dynamical constraints confine such clusters to a bounded and
narrow parameter space with effective radii between 1 pc and 15 pc,
corresponding to masses between $ \sim 10 -10^4 M_\odot$ and $\sim
10^4-10^6M_\odot$, respectively.
 We argue that gravitational microlensing is the only method
capable of constraining the abundance of dark matter in the form of
RAMBOs.

\noindent{\it Subject headings:\ }{Dark matter, Globular clusters: general,
Galaxy: kinematics and dynamics - halo, Stars: low mass, brown dwarfs}

\vfil\eject

\noindent{\bf 1. INTRODUCTION}

Observational and theoretical evidence is accumulating which suggests that dark
matter may be baryonic (see Silk 1994 for a review). Within the context of
$\Omega=1$ inflationary cosmology, the entire halos of galaxies such as our own
could consist of baryonic matter to $\simgt 50$ kpc without violating
primordial nucleosynthesis constraints. In the absence of a significant gaseous
component ({\it e.g.} Moore \& Davis 1994 and references within), such a
baryonic component is most likely to be in the form of faint stars.  The
detection of gravitational microlensing from compact objects in the galactic
halo has provided tantalising evidence that low mass stars may constitute a
significant fraction of the halo mass to distances of $\sim 30$ kpc (Alcock \et
1993, Aubourg \et 1993). A plausible scenario for the formation of these stars
is within massive clusters of similar parameters to the observed globular
clusters in our halo, which collapsed at an early epoch when Jeans mass
fluctuations in the density field became non-linear  ((Dicke
\& Peebles 1968; Fall \&
Rees 1985; Ashman 1990). The Jeans mass can be
written as $M_J \approx 2\times 10^4 M_\odot \Omega_\circ^{-1/2} h^{-3}$
($\Omega_\circ$ is the density parameter and $h$ is the Hubble
constant in units of 100 \kms\Mpc$^{-1}$), and provides a plausible guess for
the expected mass range of any large population of dark clusters that may
constitute a significant fraction of the halo mass.

It has recently been shown that Jeans mass black holes canot make up the halo
dark matter
 (Moore 1993; Rix \& Lake 1993). On the other hand, Jeans mass dark clusters
provide  an alternative dark matter candidate that can also account for the
heating of the old disk stars without producing an overly massive central bulge
as a consequence of dynamical friction, the clusters being collisionally
disrupted in the inner galaxy (Carr \& Lacey 1987).
Wasserman \& Salpeter (1994) recently proposed a scenario in which
$10^7M_\odot$
dark clusters make up 10\% of the halo mass, and contain a significant fraction
of neutron stars as well as low mass debris, mergers of which
 could provide an explanation for the frequency of
gamma ray bursts.
Gravitational microlensing searches are of particular interest in detecting
evidence for dark clusters (Maoz 1993). However, precise predictions depend on
the mass and size range adopted for the dark clusters. We shall show  in this
{\it Letter} that these parameters are tightly constrained by means of
dynamical considerations of disk heating, cluster-cluster collisions, tidal
heating of the clusters, and disruption of globular clusters.
These four constraints lead to a bounded
region in the mass \--- radius plane where dark clusters can survive intact to
the present day. We shall refer to these surviving dark clusters  as robust
associations of massive baryonic objects (RAMBOs), to distinguish them from the
unclustered population of massive compact (baryonic) halo objects (MACHOs).

RAMBOs would most probably consist of either low mass stars such as brown
dwarfs with mass in the range $0.001M_\odot-0.08M_\odot$, or faint evolved
white dwarfs in the mass range $0.4M_\odot-1.4M_\odot$.
 An unclustered
population of white dwarfs has been proposed as possible halo
dark matter (Ryu \et 1990, Tamanaha \et 1990). Current observational searches
constrain the abundance of main sequence M dwarfs ({\it e.g.} Richer \& Fahlman
1992) and of brown dwarfs (Hu \et 1994). Detecting the closest stellar dark
matter candidates expected in such scenarios is within reach of current
instrumentation. However, if these objects are clustered, then the nearest
candidates would lie much further away. The possibility of detecting both
clustered and unclustered brown dwarfs with current and future infrared
telescopes has recently been studied in detail by Kerins \& Carr (1994). In
light of our new
dynamical constraints,  we also re-examine the prospects for direct detection
via deep infrared  and CCD surveys, and  discuss the role of gravitational
microlensing searches.

\vskip 0.5truecm
\noindent{\bf 2. DYNAMICAL CONSTRAINTS ON DARK CLUSTER PROPERTIES}

\noindent{\bf 2.1 Evaporation \ \ \ }
All star clusters evaporate in a finite time due to relaxation via star-star
encounters. The two-body relaxation time of a cluster
of mass $M_{clus}$ can be written
$$
\eqalignno{
t_{rel} &= {{6.5\times 10^8{\rm yr}}\over{{\rm ln}(0.4N)}}
\Biggl({{M_{clus}}\over{10^5M_\odot}}\Biggr)^{1/2}
\Biggl({{1M_\odot}\over{m_*}}\Biggr)
\Biggl({{r_h}\over{1{\rm pc}}}\Biggr)^{3/2} \ ,
&(1)\cr }
$$
where $N$ is the number of stars of individual mass $m_*$ and $r_h$ is the
median or half--mass radius of the RAMBO of mass $M_{clus}$ (Spitzer \& Hart
1971). The evaporation timescale, $t_{evap}\approx 100t_{rel}$ (Spitzer 1975),
therefore clusters of white dwarfs with a mean radius of a parsec and of mass
equal to the smallest Jeans mass, would evaporate on a timescale of order 10
Gyrs, {\it i.e.} the Hubble time, $\tau_{_H}$ for h=1. Brown dwarf clusters of
the same mass and with $m_*=0.01M_\odot$, would survive for over two orders of
magnitude longer. In Figure 1, we show this constraint for clusters of brown
dwarfs of mass $0.02M_\odot$ and white dwarfs of mass $0.5M_\odot$.

\noindent{\bf 2.2 Tidal radii\ \ \ }
The gravitational field of the Galaxy imposes a limiting radius to a stellar
cluster. If a star passes beyond this radius, it will become unbound and escape
the cluster potential. If we ignore the potential difference across the star
cluster, then an approximate calculation yields an expression for the tidal
radius;
$$
\eqalignno{
R_{_T} &= R_{_G} \Biggl({{M_{clus}}\over{3M_{_G}}}\Biggr)^{1/3} \ ,
&(2)\cr }
$$
where $R_{_G}$ is the perigalactic distance of the star cluster and $M_{_G}$ is
the mass of the Galaxy within $R_{_G}$ ({\it e.g.} Binney \& Tremaine 1987).
When the tidal radius approaches the half mass radii, $r_h$, of the star
cluster, the rate of evaporation increases rapidly and the cluster dissolves
into the deeper potential. Stellar systems with $R_{_T}/r_h \simlt 3$ will
evaporate within $\tau_{_H}$ (Chernoff \et 1986, Oh \et 1994). The tidal radius
increases slowly with galactocentric distance and we plot this constraint for
clusters at the solar distance $R_{G,\odot}$ and $2R_{G,\odot}$.

\noindent{\bf 2.3 Cluster-cluster collisions \ \ \ }
Spitzer (1958) calculated the heating effect of giant molecular clouds upon
star clusters in the disk. We can use the same arguments to calculate the
disruption timescale of a dark cluster due to many encounters with similar
clusters. Applying the impulse approximation to a population of clusters moving
on isotropic orbits with velocity dispersion $\sigma$ yields a disruption
timescale
$$
\eqalignno{
t_{_{CC}} &= E_{bind}/{\dot E} \approx
\Biggl({{0.03\sigma}\over{GM_{clus}r_hn}} \Biggr) \ ,
&(3)\cr }
$$
where $E_{bind}\approx 0.2GM_{clus}^2/r_h$ is the cluster binding energy and
${\dot E}$ is the heating rate ({\it e.g.} Binney and Tremaine 1987). The
number density $n$ of dark clusters at large galactocentric distances $R_g$
within an isothermal halo is
$$
\eqalignno{
n &\approx \Biggl({{10^7M_\odot}\over{M_{clus}}}\Biggr)
\Biggl({{10 {\rm \ kpc}}\over{R_{_G}}} \Biggr)^2 \ {\rm kpc^{-3}} \ .
&(4)\cr }
$$
The cluster-cluster disruption timescale is very sensitive to the
galactocentric distance since $t_{_{CC}} \propto R_{_G}^2/r_h$. Adopting
$\sigma = \sqrt{(3/2)}\ 220$ \kms, we find that this constraint gives the
horizontal dashed lines drawn in Figure 1 for cluster populations at
$R_{G,\odot}$ and $2R_{G,\odot}$.

\noindent{\bf 2.4 Globular cluster heating\ \ \ }
In an identical fashion to cluster-cluster disruption, a population of massive
halo objects will inject energy into the halo globular clusters. The fact that
these systems are very old and appear spherically symmetric and undisturbed,
suggests that they are not suffering violent encounters with dark clusters. The
lack of a correlation of globular cluster properties, such as luminosity or
concentration with galactocentric radii within the Milky Way and other galaxies
(Harris 1991), also suggests that these systems are not being disrupted because
the disruption timescale $t_{_{GC}} \propto R_{_G}^2$.

The heating rate of the halo globular clusters by massive black holes was
recently calculated in detail by Moore (1993). However, several effects combine
to reduce the heating rate when the perturbers are extended systems. During a
direct collision between a dark cluster and a globular cluster with similar
mean radii, the heating rate is reduced by about an order of magnitude over the
heating from an encounter with a point-like object such as a black hole.
Consequently, black holes inject considerably more energy through penetrating
encounters than via encounters beyond the cluster tidal radius, whereas
extended perturbers inject roughly equal amounts of energy via direct
collisions as via distant collisions. Furthermore, the heating rate scales
roughly as the square of the half--mass radius of the perturber. Hence if the
dark clusters are both massive and extended, they might not have a noticeable
effect upon the Galaxy's globular clusters.

Binney \& Tremaine (1987), after Spitzer (1958), derive an analytic expression
for the heating rate of open clusters by giant molecular clouds. This formula
can be applied directly to the heating of globular clusters by dark clusters
which gives
$$
\eqalignno{
t_{_{GC}} & \approx {{0.03\sigma}\over{G}}
\Biggl({ { M_{_{GC}}}\over{r^3_{h,_{GC}}} } \Biggr)
\Biggl({{r^2_{h,clus}}\over{M^2_{clus}n}} \Biggr) \ .
&(5)\cr }
$$
For example, Palomar 5 is at a distance $R_{_G}\sim 16$ kpc and has a core
radius, $r_c \sim r_h \approx 14$ pc, and mass $M_{_{GC}} \approx 1.5\times
10^4M_\odot$. If the halo dark matter were to consist of $10^6M_\odot$ dark
clusters with effective radii of 20 pc, then Palomar 5 would be violently
heated to disruption within 1 Gyr. Although the halo globular clusters are
approximately 15 Gyr old, we shall derive our constraint on the size and radius
of dark clusters using a disruption timescale $t_{_{GC}} = 1$ Gyr. This allows
for a slow and non-violent heating of the globular clusters that otherwise
might not be observationally apparent within the present day globular cluster
population. As the cluster radii decrease, penetrating encounters begin to
inject a significant amount of energy and the globular clusters will be
disrupted on a shorter timescale. For this reason, the constraints on dark
clusters are somewhat stronger when their radii are less than that of the
observed globular clusters, and the constraint tends towards that calculated
for black holes.

The velocity dispersion of disk stars is observed to vary as the square root of
their age, a correlation which can be obtained from heating by massive black
holes (LO). To obtain the required disk heating from extended objectss, Carr \&
Lacey (1987) note that $r_h/(1 \ {\rm pc}) \simlt M_{clus} / (2\times
10^6M_\odot)$. Such clusters would have a devastating effect upon the halo
globular clusters, heating many to disruption within a tenth of a Gyr, a
timescale of order the crossing time for globular clusters such as Palomar 5
whose central velocity dispersion is $\sim 1$ \kms.

Rather than apply constraints from a specific sub-sample of low density
globular
clusters, we can use the fact that the disruption timescale varies as
$1/R_{_G}^2$, and look at the variation in the globular cluster luminosity
function with galacto-centric distance. We shall quantify this variation by
counting the fraction $f_{-7}$ of globular clusters brighter than $M_v=-7$
within inner and outer zones of the Milky Way. This magnitude corresponds to a
cluster mass of order $10^5M_\odot$ for a mass to light ratio of 1.5. We find
that for $R_{_G}<10$ kpc this fraction is 0.4, and for $R_{_G}>10$ kpc the
fraction of low mass clusters increases by less than 20\% to 0.5. The typical
poisson errors on these numbers are $\simlt 10\%$. The fractional differences
in
the numbers of low mass clusters within M31 and Virgo cluster ellipticals is
even smaller than for the Milky Way. (Note that the small difference between
the
fraction of low mass clusters can be attributed to galactic disruption
mechanisms; tidal forces, disk crossing etc., which are more effective in the
inner halo.)

The mean disruption timescale from RAMBOs is over an order of magnitude smaller
for the sample of distant globular clusters which have a mean distance of 4
times the inner sample. We can therefore constrain the properties of RAMBOs by
requiring that at $R_{_G}=10$ kpc, less than 20\% of the least massive globular
cluster have been disrupted by the present day. We find that within the
lifetime
of the observed cluster population $\sim 15$ Gyrs, the least massive 20\% of
the
globular clusters ({\it i.e.} those with mass $M_{_{GC}}< 5\times
10^4M_\odot$),
would have been disrupted by the present day if $10^5M_\odot$
RAMBOs with radius 10 pc constituted the dark matter. This constraint is in
fact
slightly stronger than we achieved above
by considering the disruption timescales of
the Palomar clusters\footnote{$^{\bf 1}$}{Applying this constraint to M31,
which has a
higher dark matter density than the Galaxy and values of
$f_{-7}$ and $f_{-8}$ which
vary by less than 10\%, yields the constraint that the halo cannot consist of
black holes of mass $M_{_{BH}} > 10^3M_\odot$.}.
(For these calculations we took
the core radius of the dark matter distribution to be 10 kpc.)
Applying this result to the model proposed by Wasserman \& Salpeter (1994),
we find that a fraction $<1\%$ of the galactic halo could
be composed of RAMBOs of mass $10^7M_\odot$ with internal dispersions
$\sim 100$ \kms, an order of magnitude lower than they proposed.
 We therefore conclude that
the observed correlation between age and velocity dispersion of disk stars
cannot arise from heating by either massive compact halo dark matter
objects or by RAMBOs.


\vskip 0.5truecm
\noindent{\bf 3. OBSERVATIONAL PROSPECTS FOR DETECTING DARK CLUSTERS}
\vskip 0.5truecm

\noindent{\bf 3.1 Infrared Observations \ \ \ }
If the dark matter were clustered, then this would have important implications
for the direct detection experiments currently in progress ({\it e.g.} Hu \et
1994). The distance to the nearest cluster will be larger by a factor
$(M_{clus}/m_*)^{1/3}$, but the luminosity is increased by the larger factor
$(M_{clus}/m_*)$. This increase in flux is only useful if the clusters can be
treated as point sources. For a random distribution, the closest cluster will
lie at a distance $\approx 0.5 n^{-1/3} \equiv 300 M_6^{1/3}$ pc where
$M_6=M_{clus}/10^6M_\odot$, and the angular size of this cluster will be
$\approx 2r_{10}M_6^{-1/3}$ degrees on the sky, where $r_{10} = r_h/10 {\rm \
pc}$. The closest $10M_\odot$ brown dwarf cluster would therefore lie about 6
pc from the sun and subtend over ten degrees on the sky. Thus a full sky survey
would be necessary in order to detect the nearest halo brown dwarf RAMBO. Both
of the future infrared observatories, ISO and SIRTF, are planned to take
observations only in pointed mode.

Essentially all of the clusters which lie within the dynamically allowed region
in Figure 1 can be treated as extended sources, {\it i.e.} a cluster with $r_h$
of $\sim 1 $ pc at 50 kpc subtends several arcseconds on the sky. The regime in
which every line of sight towards the LMC contains at least one RAMBO is
indicated on Figure 1 and this bisects the allowed parameter space. To the
right of this line, the clusters will give rise to a general galactic
background with Poisson fluctuations in intensity. To the left of this line, a
certain fraction of the sky must be observed in order to find one or more
clusters along the line of sight.

The region probed by IRAS and ISO lies to the right of the allowed region
indicated in Figure 1 ({\it c.f.} Kerins \& Carr 1994). The detection criterion
was based on brown dwarfs of mass $m_*=0.02M_\odot$, using an extended source
sensitivity at $6.75\mu{\rm m}$ for ISO for a $3\sigma$ detection from a 2 day
observation. These calculations were based on the assumption that the brown
dwarfs are blackbody radiators. Recent calculations by Saumon \et (1994) which
include updated opacities show that the emergent spectra do not resemble
blackbody emission; however the location in the colour-magnitude diagram of the
low mass stars does not change significantly. Brown dwarfs of the maximum
allowed mass, $m_*=0.08M_\odot$, would have an intensity an order of magnitude
higher, although the evaporation constraints are correspondingly larger and the
extended emission from these clusters would still not be visible by ISO. Note
that  Rix \& Lake (1993) conclude that RAMBOs of mass $2\times 10^6M_\odot$
would have been detected within the IRAS point source survey. However, this
result is only correct if the nearest clusters appear as point sources, {\it
i.e.} $r_h \simlt 0.01$ pc, a scale much smaller than the clusters considered
by these authors, and clusters of this size are already ruled out by
evaporation
constraints.


\noindent{\bf 3.2 CCD Imaging \ \ \ }
An unclustered distribution of halo white dwarfs is detectable with current
instrumentation. Within 10 pc of the sun, we expect of order 40 white dwarfs
from the halo, the closest lying about 3 pc away. For a halo age of 15 Gyr, an
individual white dwarf luminosity would be $\sim 10^{-5.5}L_\odot$ and the
closest white dwarf would have an apparent magnitude $m_{_B}=16.6$. The use of
proper motions to distinguish between disk and halo stars will enable current
searches to constrain the number of unclustered halo white dwarfs. However if
the white dwarfs are clustered, then the detection of these objects becomes
significantly harder. The nearest white dwarf RAMBO would have a minimum mass
$M_{clus}=200M_\odot$ and corresponding radius $r_h\sim 4$ pc, and would
therefore lie at about 18 pc and have an angular size of 13 degrees. Individual
stars within the cluster would have $m_{_B}\approx 21$. Distinguishing between
disk and halo white dwarfs would be extremely difficult and require extensive
proper motion studies.

Observational searches are therefore limited by the low surface brightness of
the clusters. One might expect to obtain images with sensitivity at best
$\mu_{_B} = 29$ mag arcsec$^{-2}$, similar to the deepest faint galaxy count
surveys. Let us consider an optimum case, a white dwarf cluster of
$10^4M_\odot$ and $r_h\approx 1$ pc. This object would have a total magnitude
$M_{_B}\sim 9$, or $m_{_B} \sim 24$ at 10 kpc. However, at this distance the
cluster would cover over 1000 arcsec$^2$ and have a surface brightness of
$\mu_{_B}\approx 32$ mag arcsec$^{-2}$. We have plotted on Figure~1 the region
currently accessible by deep CCD searches which reach a surface brightness
$\mu_{_B} = 29$ mag arcsec$^{-2}$. Very careful observations may be able to
probe a small part of the allowed parameter space, especially for a halo age
less than 15 Gyr, however the sky background and readout noise will prohibit a
thorough investigation of the entire allowed region.

Could RAMBOs consisting of low mass stars at the edge of the main sequence
remain undetected by deep CCD searches? Consider as above, a $10^4M_\odot$
cluster of zero metallicity stars with $m_*\sim 0.1M_\odot$. Individual stars
have absolute red magnitudes $M_{_R}=12$ (Burrows \et 1994), therefore the
absolute magnitude of this cluster would be $M_{_R}=-0.5$. At 10 kpc, its
apparent magnitude $m_{_R}=14.5$ gives rise to a surface brightness $\mu_{_R}
\approx 22$ mag arcsec$^{-2}$. Deep searches can reach $R\approx 27$, and such
clusters would  be detectable because of the intense diffuse light  background
they produce. However, if the metal abundances of the same stars were equal to
the solar value, then $M_{_R}=18$, and the surface brightness of the cluster
would be a magnitude fainter than the current detection threshold. An
observational sensitivity at the $K$ band surface brightness prdedicted for
this cluster, $\mu_{_K}=21$ mag arcsec$^{-2}$, has been reached on the Keck
telescope (Hu \et 1994).

\noindent{\bf 3.3 Gravitational Microlensing \ \ \ }
Infrared and CCD observations appear to be incapable of probing much, if any,
of the allowed parameter space for dark star clusters. Gravitational
microlensing is proving to be an interesting technique for determining the
stellar content of dark matter halos. Although the total extent of galaxy halos
is still undetermined, observational evidence suggests that the halo of the
Milky Way extends to beyond the Large Magellanic Cloud (LMC) and is at least
$10^{12}M_\odot$ (Fich \& Tremaine 1992). Lensing experiments towards the LMC
are typically sensitive only to material between the sun and typically to half
of the distance to the source, {\it i.e.} a total of $5\times 10^6M_\odot$ of
dark matter per square degree along the line of sight to the LMC. Most of the
stars lie within 10 square degrees, although to date, only a couple of square
degrees have been searched.

These statistics are interesting because if the dark matter consists of RAMBOs,
the lensing results may be dominated by small number statistics since only a
few dark clusters may currently lie in the field of view. For example, if the
RAMBOs have  mass $3 \times 10^6M_\odot$, then the searches of Alcock \et may
only have $3\pm\sqrt{3}$ clusters within the field of view. As long as most of
the LMC is monitored, microlensing will be able to provide constraints on dark
matter in the form of star clusters. The most massive clusters are constrained
to have half--mass radii of order 50 pc, and they have angular sizes of half of
a degree at 10 kpc from the sun. The smallest dark cluster, $r_h \sim 1$ pc,
would have an angular size of about an arcminute, and although the microlensing
optical depth would be larger within this area, the total number of stars which
could be lensed is down by the same factor and the lensing event rate is
identical to that of an unclustered population of stars.

All the events from one cluster would have the same net motion of the cluster,
with the internal velocity dispersion superimposed, $\sigma \approx
\sqrt{GM_{clus}/4r_{clus}}$, which  at maximum is $\sim 10$ \kms, although
within most of the dynamically allowed parameter space, clusters would have
internal dispersions of order $\sigma \approx 1$ \kms. The lensing events from
a single cluster will yield information on the mass function of the stars
rather than the internal velocity dispersion of the cluster.  Recovering $r_h$
and the masses of individual RAMBOs will require $\simgt 10$ events per cluster
and is complicated by the fact that the covering factor of these objects is
close to unity. Figure 2 shows the expected distribution of RAMBOs projected
onto a region of the sky 3 deg$^2$, which is about equal to the maximum extent
of the LMC which can be monitered for lensing events. For this diagram we have
assumed that the RAMBOs have mass $10^6M_\odot$ and size 60 pc. This yields a
number, $N_{_{LOS}}$, of order one cluster for every line of sight towards the
LMC, and therefore for a Poisson distribution the fraction of sky actually
covered is $1-{\rm exp}(-N_{_{LOS}})=63$\%.

Recent preprints by Maoz (1994) and Bouquet \et (1994) discuss the possibility
of detecting the microlensing signatures of dark star clusters of mass
$10^6M_\odot$ and $r_h\approx 1$ pc as proposed by Carr \& Lacey (1987). We
have shown that these clusters are excluded by dynamical constraints.
Furthermore, the allowed parameter space for RAMBOs (Figure~2) indicates that
these
systems have a large covering factor and small internal velocity dispersions,
and therefore their microlensing signatures will be very difficult to measure.

{\noindent This research has been supported at Berkeley in part by grants from
the N.S.F.}

\noindent{\bf References}
\parindent=0pt

\pp Alcock, C. \et 1993, {\it Nature}, {\bf 365}, (no. 6447) 623.

\pp Ashman, K.M. 1990, {\it M.N.R.A.S.}, {\bf 247}, 662.

\pp Aubourg, E. \et 1993, {\it Nature}, {\bf 365}, (no. 6447) 621.

\pp Binney, J. and Tremaine, S. 1987, {\it Galactic Dynamics}, {\it Princeton
University Press,} ed. J. Ostriker.

\pp Bouquet, A., Kaplan, J. \& Verschaeve, J.M. 1994, {\it astro-ph \#9402029}.

\pp Burrows, A., Hubbard, W.B. \& Lunine, J.I. 1994, {\it Eighth Cambridge
Workshop on Cool Stars, Stellar Systems and the Sun}, in press.

\pp Carr, B.J. 1978, {\it Comm.Astrophys.}, {\bf 7}, 161.

\pp Carr, B.J. and Lacey, C.G. 1987, {\it Ap.J.}, {\bf 316}, 23.

\pp Chernoff, D.F., Kochanek, C.S. and Shapiro, S.L. 1986, {\it Ap.J.}, {\bf
309}, 183.

\pp Dicke, R.H. and Peebles, P.J.E. 1968, {\it Ap.J.}, {\bf 154}, 891.

\pp Fall, S.M. \& Rees, M.J. 1985, {\it Ap.J.}, {\bf 298}, 18.

\pp Fich, M. and Tremaine, S. 1992, {\it Ann.Rev.Astr.Astro.},{\bf 29}, 409.

\pp Harris, W.E. 1991, {\it A.R.A.A.}, {\bf 29}, 543.

\pp Hu, E.M., Huang, J.-S., Gilmore, G. \& Cowie, L.L. 1994,
Institute of Hawaii preprint.

\pp Hut, P. \& Rees, M.J. 1992, {\it M.N.R.A.S.}, {\bf 259}, 27P.

\pp Kerins, E.J. \& Carr, B.J. 1994, {\it M.N.R.A.S.}, in press.

\pp Lacey, C.G. and Ostriker, J.P. 1985, {\it Ap.J.}, {\bf 299}, 633.

\pp Maoz, E. 1994, {\it astro-ph \#9402027}.

\pp Moore, B. 1993, {\it Ap.J.Lett.}, {\bf 413}, L93.

\pp Moore, B. \& Davis, M. 1994, {\it M.N.R.A.S.}, in press.

\pp Oh, K.S., Lin, D.N.C. \& Aarseth, S.J. 1994, {\it M.N.R.A.S.}, submitted.

\pp Richer, H.B. \& Fahlman, G.G. 1992, {\it Nature}, {\bf 358}, 383.

\pp Rix, H.W. \& Lake, G. 1993, {\it Ap.J.Lett.}, {\bf 417}, L1.

\pp Ryu D, Olive K.A. \& Silk J. 1990, {\it Ap.J.}, {\bf 353}, 81.

\pp Silk, J. 1994, in  {\it Cosmology and Large--Scale Structure},
 proceedings of the 1993 Les Houches Summer School on Theoretical Physics,
eds. R. Schaeffer, M. Spiro and J. Silk (Dordrecht: Elsevier), in press.

\pp Spitzer, L. 1958, {\it Ap.J.}, {\bf 127}, 17.

\pp Spitzer, L. \& Hart, M.H. 1971, {\it Ap.J.}, {\bf 164}, 399.

\pp Spitzer, L. 1975, {\it Dynamics of Stellar Systems}, I.A.U. 69.
ed. A.Hayli, (Dordrecht: Reidel).

\pp Tamanaha C.M., Silk J., Wood M.A. \& Winget D.E. 1990, {\it Ap.J.}, {\bf
358}, 164.

\pp Wasserman I. \& Salpeter E.E. 1994, Cornell preprint no. 1049.

\pp Wielen, R. 1985, {\it I.A.U. Symp.}, {\bf 113}, 449.

\vskip 1.0truecm
\parindent=36pt
\noindent{\bf Figure Captions}

{\noindent Figure 1.} \ \ The mass \--- radius parameter space within which
RAMBOs can survive until the present day.  Regions at the text side of the
curves are excluded by the four dynamical constraints discussed in Section 2:
evaporation, tidal disruption, cluster-cluster disruption and globular cluster
heating are indicated by the solid, dot-dashed, dashed and dotted lines
respectively. Cluster-cluster disruption and tidal disruption are sensitive to
$R_{_G}$ and constraints are shown for clusters at the solar radius
$R_{G,\odot}$, and $2R_{G,\odot}$. The evaporation constraint is shown for
brown dwarf clusters with $m_*=0.02M_\odot$, and white dwarf clusters with
$m_*=0.5M_\odot$. The region where the number of clusters per line of sight is
larger than unity is indicated and bisects the allowed parameter space.
Clusters to the right of this region give rise to a galactic background with
Poisson fluctuations in intensity. IRAS and ISO extended sky surveys can only
detect brown dwarf clusters clusters to the right of the indicated lines.
Similarly, deep CCD observations can only detect clusters of white dwarfs
outside of the dynamically allowed region.

\vskip 1.0truecm
{\noindent Figure 2.} \ \ A random realisation of the distribution of RAMBOs
projected onto a 3 deg$^2$ region of the sky, similar to the maximum extent of
the LMC. We have assumed a number density of RAMBOs given by equation (4) with
individual mass $M_{clus}=10^6M_\odot$ and $r_h=60$ pc. Each line of sight is
expected to contain a single RAMBO on average, therefore the covering factor
here is 0.63.

\bye